\documentclass[fleqn,usenatbib]{mnras}

\usepackage[T1]{fontenc}
\usepackage{ae,aecompl}

\usepackage[colorinlistoftodos]{todonotes}

\usepackage{graphicx}	
\usepackage{amsmath}	
\usepackage{amssymb}	
\usepackage{media9}
\usepackage{siunitx}

\definecolor{MMgreen}{RGB}{0,128,0}

\def\mso{{\rm M}_\odot}

\def\rso{{\rm R}_\odot}
\def\rst{{\rm R}_\star}
\def\lso{{\rm L}_\odot}

\def\kms{{\rm km}\,{\rm s}^{-1}}

\def\msoy{\mso\,{\rm yr}^{-1}}

\newcommand{\tento}[1]{\times 10^{#1}}

\title[CO $J = 2-1$ NOEMA observations of $\mu$ Cep]{NOEMA maps the CO $J = 2-1$ environment of the red supergiant $\mu$ Cep\thanks{Cleaned images as FITS files and basic stellar parameters are available at the CDS (\url{http://vizier.u-strasbg.fr/viz-bin/VizieR}).}}

\author[M. Montarg\`es et al.]{
M.~Montarg\`es$^{1,}$\thanks{E-mail: miguel.montarges@kuleuven.be},
W.~Homan$^{1}$,
D.~Keller$^{1}$,
N.~Clementel$^{1}$,
S.~Shetye$^{2,1}$,
L.~Decin$^{1}$,
\newauthor
G.~M.~Harper$^{3}$,
P.~Royer$^{1}$,
J.~M.~Winters$^{4}$,
T.~Le~Bertre$^{5, 6}$,
and A.~M.~S.~Richards$^{7}$.
\\
$^{1}$Institute of Astronomy, KU Leuven, Celestijnenlaan 200D B2401, 3001 Leuven, Belgium\\
$^{2}$Institute of Astronomy and Astrophysics, Universit\'e libre de Bruxelles, B-1050  Bruxelles, Belgium\\
$^{3}$Center for Astrophysics and Space Astronomy, University of Colorado, Boulder, CO 80309, USA\\
$^{4}$Institut de Radioastronomie Millim\'etrique, 300 rue de la Piscine, 38406, Saint Martin d'H\`eres, France\\
$^{5}$LERMA, Observatoire de Paris, PSL Research University, CNRS, UMR 8112, 75014, Paris, France\\
$^{6}$Sorbonne Universit\'es, UPMC Univ. Paris 06, UMR 8112, LERMA, 75005, Paris, France\\
$^{7}$Jodrell Bank Centre for Astrophysics, School of Physics and Astronomy, University of Manchester, Manchester M13 9PL, UK
}

\date{Accepted 2019 February 04. Received 2019 January 26 ; in original form 2018 December 18}

\pubyear{2019}

\begin{document}
\label{firstpage}
\pagerange{\pageref{firstpage}--\pageref{lastpage}}
\maketitle

\begin{abstract}
Red supergiant stars are surrounded by a gaseous and dusty circumstellar environment created by their mass loss which spreads heavy elements into the interstellar medium.
The structure and the dynamics of this envelope are crucial to understand the processes driving the red supergiant mass loss and the shaping of the pre-supernova ejecta.
We have observed the emission from the CO $J = 2-1$ line from the red supergiant star $\mu$~Cep with the NOEMA interferometer. In the line the synthesized beam was $0.92 \times 0.72$~arcsec ($590 \times 462$~au at 641~pc). The continuum map shows only the unresolved contribution of the free-free emission of the star chromosphere. The continuum-subtracted channel maps reveal a very inhomogeneous and clumpy circumstellar environment. In particular, we detected a bright CO clump, as bright as the central source in the line, at 1.80~arcsec south-west from the star, in the blue channel maps. After a deprojection of the radial velocity assuming two different constant wind velocities, the observations were modelled using the 3D radiative transfer code \textsc{lime} to derive the characteristics of the different structures.
We determine that the gaseous clumps observed around $\mu$~Cep are responsible for a mass loss rate of $(4.9 \pm 1.0) \tento{-7}~\msoy$, in addition to a spatially unresolved wind component with an estimated mass-loss rate of $2.0 \tento{-6}~\msoy$. Therefore, the clumps have a significant role in $\mu$~Cep's mass loss ($\ge 25 \%$). We cannot exclude that the unresolved central outflow may be made of smaller unresolved clumps. 
\end{abstract}

\begin{keywords}
circumstellar matter --  stars: imaging -- stars: individual: $\mu$~Cep -- stars:~mass-loss -- supergiants -- radio lines: stars
\end{keywords}



\section{Introduction}\label{Sect:Intro}

Cool evolved stars are among the most important contributors to the chemical evolution of the Universe. They are characterised by a circumstellar environment (CSE) created by the stellar wind. Within this outflow, atoms are forming molecules that can condense into dust. During their evolved stage, low and intermediate mass star (M $\le 8$~$\mso$) go through the asymptotic giant branch (AGB) stage. At that time, the star experiences a pulsation-enhanced dust-driven wind that is believed to generate this CSE \citep{2018A&ARv..26....1H}. More massive stars, that evolve into red supergiant (RSG) stars, exhibit a similar gaseous and dusty CSE. However, there is currently no consistent scenario to explain their mass loss (e.g., \citealt{2017A&A...602L..10O} and references therein).

$\mu$~Cep (Erakis, Herschel's Garnet Star, HR~8316, HD~206936) is an M2-Ia star, surrounded by a CSE originating from its mass loss. It has no confirmed binary companion. Estimates of $\mu$~Cep's distance vary between $390 \pm 140$ and $1818 \pm 661$~pc (\citealt{2007A&A...474..653V,2008ApJ...685L..75D,2005A&A...436..317P} and references therein). The literature gives $v_\mathrm{LSR} = 23~\kms$ for $\mu$~Cep (LSR is the Local Standard-of-Rest frame). However, this value is derived from a erroneous association to a cluster (Trumpler~37 in IC1396, \citealt{2008A&A...485..303M}). \citet{1989A&A...210..198L} derived a value of $35~\kms$ and \citet{1953GCRV..C......0W} determined a mean optical photospheric heliocentric radial velocity of $v_\mathrm{helio} = 19.3 \pm 0.5~\kms$. In App.~\ref{Sect:star_vel}, we determine the systemic velocity of $\mu$~Cep to be $v_\mathrm{LSR} = 32.7 \pm 0.1~\kms$. At 25~$\mu$m, \cite{2008ApJ...685L..75D} imaged an asymmetric dusty nebula with a size of several arcseconds, extended from the North North-East to the South-West direction. \citet{2016AJ....151...51S} observed the envelope over a broader spectral range from the mid to the far infrared. Adopting a distance of 870~pc, they modelled a dust distribution between 96 and 96\,000~au around the star and derived a mass-loss rate of $4 \tento{-6}~\msoy$, with evidence for a decline over the past 13\,000 years. Much closer to the star, \citet{2000ApJ...538..801T} and \citet{2005A&A...436..317P} characterized the inner molecular envelope (MOLsphere) surrounding the star, identifying CO and H$_2$O at less than 0.5 stellar radius from the photosphere. $\mu$~Cep has a very peculiar non-parabolic CO $J=2-1$ profile \citep{1989A&A...210..198L}. The CO molecule has a low dipole moment \citep[][and references therein]{2016CPL...663...84C}, which makes it relatively insensitive to the surrounding radiation field. Infrared pumping of ground or vibrationally excited states of CO is expected to be minimal \citep{2019A&A...622A.123D}. Therefore, its main excitation mechanism is through collisions which implies that it is a good tracer of density (when optically thin) and temperature (when optically thick). Therefore, it is an excellent molecule to investigate circumstellar morphology. From the single-dish observations of several CO rotationally excited lines ($J = 3-2$ and $J = 4-3$), \citet{2010A&A...523A..18D} derived a mass loss rate of $\sim 2 \tento{-6}~\msoy$, however this value is based on a previously determined distance of 390~pc and an incorrect systemic velocity ($v_\mathrm{LSR} = 23~\kms$). 

By obtaining mm-interferometry observations of $\mu$~Cep, we intend to image its CSE to determine how the stellar mass loss is shaping the gaseous environment. We present the observations and data reduction in Sect.~\ref{Sect:Obs}. We proceed with a description of the intensity maps and integrated spectrum, and also perform a deprojection of the velocity cube in Sect.~\ref{Sect:Morpho}. We present the results of the modelling of the CO emission through three-dimensional radiative transfer in Sect.~\ref{Sect:MassLoss}. We discuss the wind velocity field model as well as the mass-loss rate and mechanisms in Sect.~\ref{Sect:Discussion}. Finally, our concluding remarks are presented in Sect.~\ref{Sect:Conclusion}.

\section{Observations and data reduction}\label{Sect:Obs}

$\mu$~Cep was observed with NOEMA (NOrthern Extended Millimeter Array) on 2015 December 2 and 2016 March 24 in the 7C (3.4h on source) and 7B (2.6h on source) configurations, respectively. The baseline lengths ranged from 20 to 192~m, and from 42.3 to 452~m, respectively. The average system temperature and precipitable water vapour were 120~K and 1.5~mm, respectively, during both observation runs. The line data were provided through the narrow band backend using the 160~MHz bandwidth units on both polarizations. We obtain a spectral resolution of $0.81~\kms$ over a range of $\pm 90~\kms$ centred on the CO $J=2-1$ line at 230.538~GHz. The Wideband Express (WideX) backend was used to produce a continuum dataset centred at 231.276~GHz, selecting the line free channels over the 4~GHz bandwidth.

The data were reduced and calibrated using \textsc{clic}, which is part of the publicly available \textsc{gildas} package\footnote{\url{http://www.iram.fr/IRAMFR/GILDAS}}. The phase centre was set at $(\alpha_{2000}, \delta_{2000}) = (21^h43^m30.461^s, \, +58\degr46\arcmin48.160\arcsec)$. A self calibration was performed on the continuum ($u,v$) data and we applied the resulting gain to the spectral data. The imaging and cleaning were done using Briggs weighting (robust set to 1). The resulting synthesized beam is $0.93 \times 0.70$~arcsec for the continuum and $0.92 \times 0.72$~arcsec for the CO line. The maximum recoverable scale is 8~arcsec. Each line spectral channel was continuum subtracted in the ($u, v$) domain. The quasars 2037+511, 2146+608, and J2201+508 were used as phase and amplitude calibrators. The absolute calibration was performed using the standard flux calibrators MWC349 and LkHa101. In the 1.3~mm band (band 3 of NOEMA), the uncertainty on the flux calibration is $\le 20\%$. The $1\sigma$ level is 0.71~mJy\,beam$^{-1}$ in the continuum map and $2.03 - 4.70$~mJy\,beam$^{-1}$m in the line channel maps.

\section{Description of the data}\label{Sect:Morpho}

\subsection{Continuum}

The self-calibrated continuum map of $\mu$~Cep, centred at 231.276~GHz, is represented on Fig.~\ref{Fig:continuum_map}. Only the central source is visible at a level higher than $3\sigma$. The continuum image shows no departure from spherical geometry. It presents a small offset of ($0.03$~arcsec,  $-0.02$~arcsec) with respect to the phase centre, both in the original and the self-calibrated maps. {The star remains unresolved in the continuum map and has a point flux density of $39.67 \pm 7.93$~mJy. (\citealt{1994A&A...281..161A} measured a flux density of $59.0 \pm 6.0$~mJy in their 250~GHz continuum survey with the IRAM 30m telescope). 

Using an updated model of \citet{2001ApJ...551.1073H}, we can derive the main source of the continuum emission if we assume that $\mu$~Cep has the same atmospherical structure as Betelgeuse, the prototypical M2I RSG. By scaling the angular diameter from $44.06 \pm 0.59$~mas for Betelgeuse \citep{2016A&A...588A.130M} to $14.11 \pm 0.60$~mas for $\mu$~Cep \citep{2005A&A...436..317P}, we derive a flux density at 231.3~GHz of $36 \pm 4$ mJy and $r(\tau = 1/3) \sim 1.3$~R$_\star$. Therefore, from this modelling we can conclude that most of the continuum emission comes from the relatively compact free-free emission in the heated extended stellar atmosphere (chromosphere). Any dust contribution would appear more diffuse and extended.

\begin{figure}
	\includegraphics[width=\columnwidth]{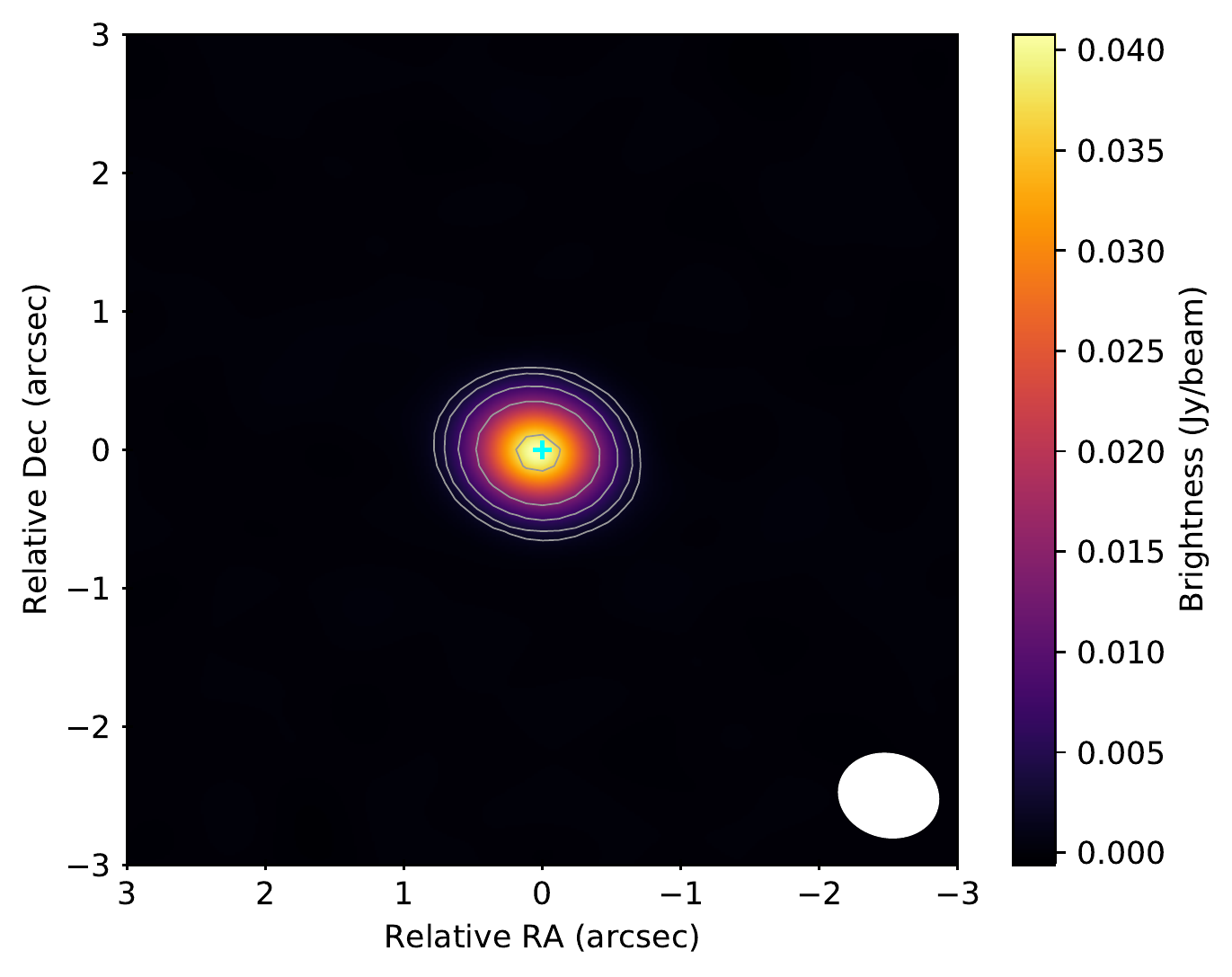}
	\caption{Self calibrated continuum map of $\mu$~Cep centred at 231.276~GHz. The synthesized beam is represented by the white ellipse at the bottom right corner of the image. The cyan cross marks the position of the star at the (0, 0) relative coordinates. The contour levels are 3, 5, 10, 20 and 50 times the noise rms ($1\sigma =  0.71$~mJy\,beam$^{-1}$ in the continuum).\label{Fig:continuum_map}}
\end{figure}
 
\subsection{Line profile\label{Sect:spectrum}}
 
The integrated line profile is represented on Fig.~\ref{Fig:spectrum}, and is very asymmetric. Moreover, the line profile of the central beam area (red curve) is not strictly double peaked. However, we can identify a red horn centred at the stellar systemic velocity. Its peak value is 147~mJy. In addition to the main red horn, the central aperture line profile presents a secondary horn at roughly $45~\kms$ at 100~mJy. There is also a blue horn centred at $-2.5~\kms$ and peaking at 147~mJy.
 
\begin{figure}
	\includegraphics[width=\columnwidth]{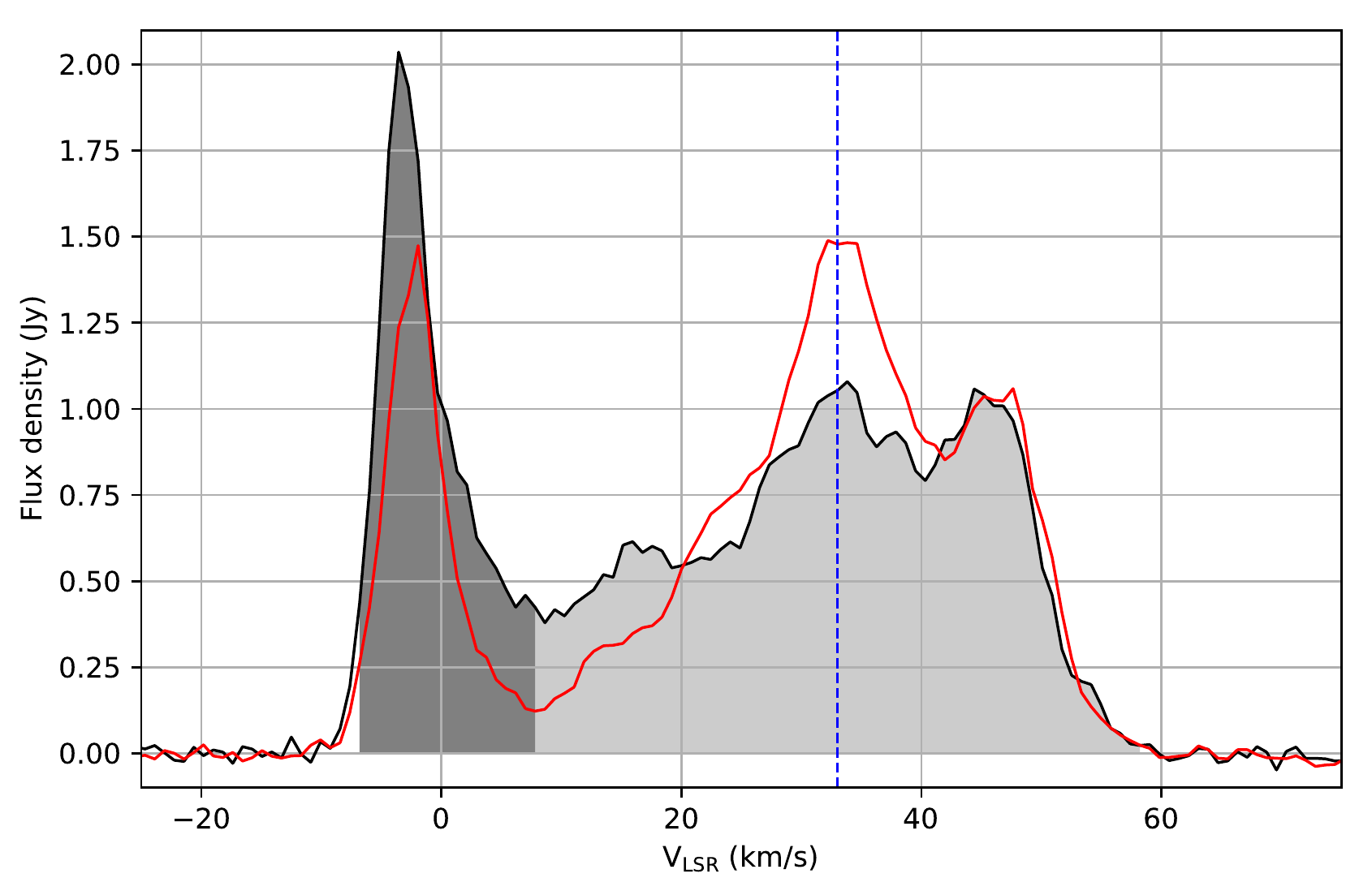}
	\caption{Integrated line profile over the central $12\times12$~arcsec region of the continuum subtracted CO $J=2-1$ (230.538~GHz) maps of $\mu$~Cep in black. The red curve correspond to 10 times the flux density integrated over a beam size aperture at the phase centre. The vertical dashed blue line corresponds to the star velocity derived in App.~\ref{Sect:star_vel}. The light grey area corresponds to the slow wind and the dark grey area to the fast wind (see Sect. \ref{Sect:Deproj}).\label{Fig:spectrum}}
\end{figure}

\subsection{Channel maps \label{Sect:channel_maps}}

Figure~\ref{Fig:channel_maps} represents the channel maps in the CO $J=2-1$ line of $\mu$~Cep after continuum subtraction. First, we notice a central bright spot in all the non-empty channels. Its intensity is highly dependent on the velocity channel. It is also the case for its position relative to the phase centre: in the red channels (v$_\mathrm{LSR} > 32.7~\kms$) it coincides with the phase centre, while in the blue channels (v$_\mathrm{LSR} < 32.7~\kms$) it is shifted to the South-West between 20.0 and 2.9~$\kms$ and to the East between $-3.6$ and $-10.1~\kms$. Though a bright central spot is expected for a smooth spherical outflow component at the higher velocities, its persistence over the whole velocity range and  its changing location are difficult to explain with this simplified spherical symmetry only. 

The channel maps reveal several small- and larger-scale clumpy features (Fig.~\ref{Fig:channel_maps}). We labelled each of them and give their positions in Table~\ref{Tab:Feature_Id}, although they could be made of smaller structures unresolved by NOEMA. The clumps are not visible in all the spectral channels and are present at various position angles (PA) and distances from the phase centre. The most prominent features are visible in the blue channels from $-8.4$ to $7.8~\kms$. The first one (C1), is located on average at 1.80~arcsec from the star (1.15~kau at 641~pc, on the plane of the sky). The second one (C2) coincides with the star at the centre of its channels but is shifted toward the West in the red, and toward the East in the blue.

\begin{figure*}
	\includegraphics[width=\textwidth]{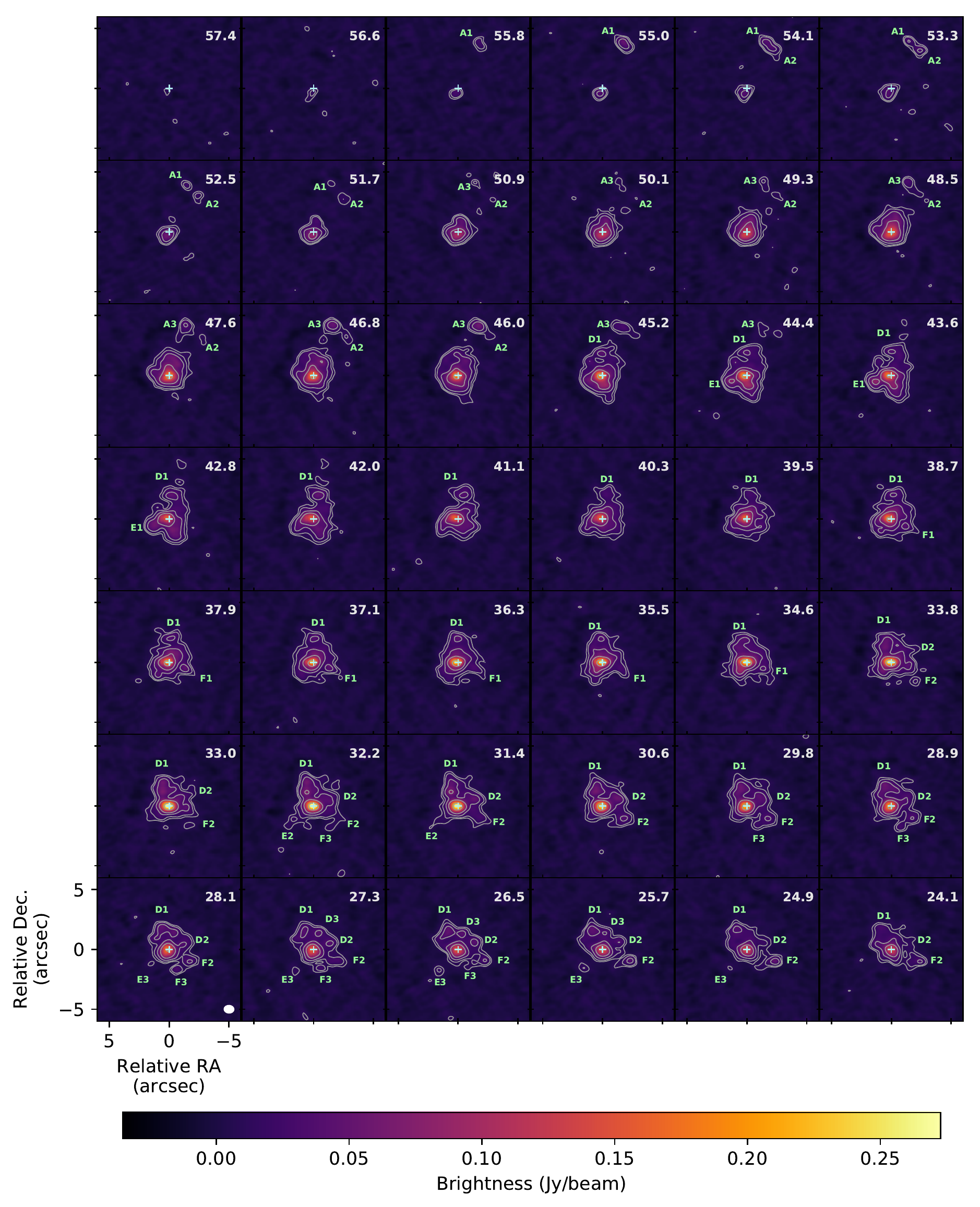}
    \caption{Continuum subtracted channel maps of $\mu$~Cep from the NOEMA observations, centred at 230.538~GHz. The representation is limited to non-empty channels (57.4 to $-10.1~\kms$) The LSRK (LSR kinematic frame) radial velocity in $\kms$ is expressed at the top right corner of each map. The LSRK velocity of the star is $32.7 \pm 0.1~\kms$ (App.~\ref{Sect:star_vel}). The synthesized beam is represented by the white ellipse at the bottom right corner of the first image of the last row. On each map, the pale blue cross marks the position of the star at the (0, 0) relative coordinates. The contour levels are 3, 5, 10, 20 and 50 times the noise rms of the respective channel ($1 \sigma = 2.03 - 4.70$~mJy\,beam$^{-1}$). The clumps are identified by the pale green labels (see details in Table \ref{Tab:Feature_Id}). \label{Fig:channel_maps}}
\end{figure*}

\begin{figure*}
	\includegraphics[width=\textwidth]{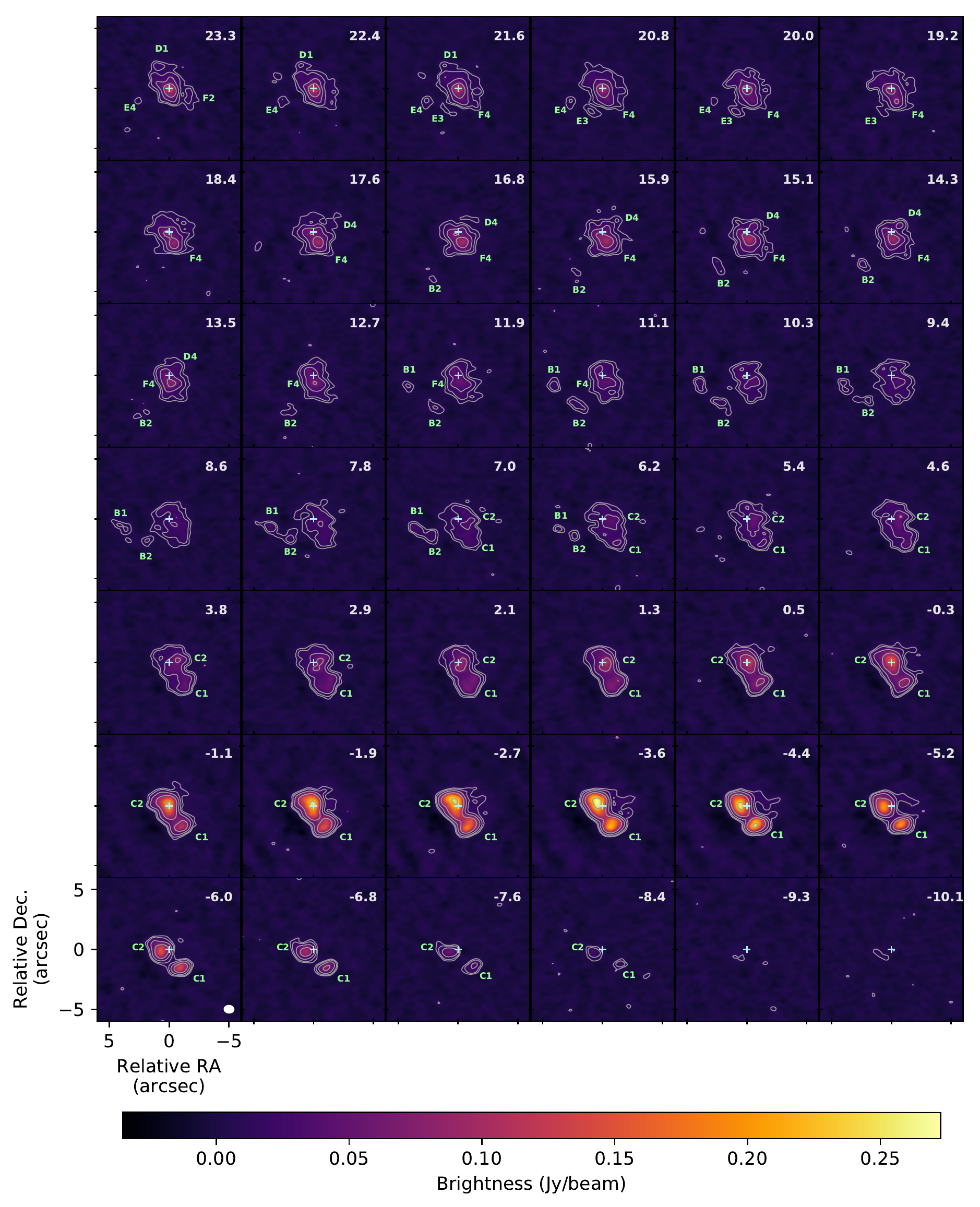}
    \contcaption{}
\end{figure*}

\begin{table}
	\centering
	\caption{Identification and qualitative positions of the clumps in the channel maps of $\mu$~Cep (Fig.~\ref{Fig:channel_maps}). $v^\mathrm{mes}_\mathrm{LSR}$ corresponds to the velocity at which the relative right ascension and declination were measured. $v^\mathrm{min - max}_\mathrm{LSR}$ is the velocity interval in which the feature is visible. The first part contains the most prominent features that are modelled in Sect.~\ref{Sect:MassLoss}. Features not incorporated in the modelling are listed in the second part of the table.}
	\sisetup{
		table-number-alignment = center,
		table-figures-integer  = 2
	}
	\label{Tab:Feature_Id}
	\begin{tabular}{c S S S c}
		\hline
		Id. & {$\Delta \alpha$} & {$\Delta \delta$} & {$v^\mathrm{mes}_\mathrm{LSR}$} & $v^\mathrm{min - max}_\mathrm{LSR}$ \\
		& {(arcsec)} & {(arcsec)} & {($\kms$)} & ($\kms$) \\
		\hline
		A1 & -1.42 & 3.92 & 52.5 & $51.7 \rightarrow 55.8$ \\
		A2 & -2.37 & 2.97 & 52.5 & $46.0 \rightarrow 54.1$\\
		A3 & -1.66 & 4.15 & 46.8 & $44.4 \rightarrow 50.9$ \\
		B1 & 3.95 & -0.89 & 9.4 & $6.2 \rightarrow 11.9$ \\
		B2 & 2.02 & -2.02 & 9.4 & $6.2 \rightarrow 16.8$ \\
		C1 & -0.71 & -1.66 & -4.4 & $-8.4 \rightarrow 7.0$ \\
		D1 & -0.47 & 2.02 & 41.1 & $21.6 \rightarrow 45.2$ \\
		F4 & -0.36 & -0.95 & 17.6 & $11.1 \rightarrow 21.6$\\
		\hline
		C2 &  0.00 & -0.12 & -1.9 & $-8.4 \rightarrow 7.8$ \\
		D2 & -1.42 & 0.47 & 31.4 & $24.1 \rightarrow 33.8$ \\
		D3 & -0.36 & 1.31 & 27.3 & $25.7 \rightarrow 27.3$ \\
		D4 & -0.71 & 0.59 & 14.3 & $13.5 \rightarrow 17.6$ \\
		E1 & 1.31 & -0.47 & 43.6 & $42.8 \rightarrow 44.4$ \\
		E2 & 1.66 & -1.07 & 32.2 & $31.4 \rightarrow 32.3$\\
		E3 & 1.54 & -1.66 & 26.5 & $19.2 \rightarrow 21.6$\\
		& & & & $25.7 \rightarrow 28.1$ \\
		E4 & 2.61 & -0.95 & 21.6 & $20.0 \rightarrow 23.3$ \\
		F1 & -1.31 & -0.47 & 37.9 & $34.6 \rightarrow 38.7$ \\
		F2 & -2.37 & -0.95 & 24.9 & $23.3 \rightarrow 33.8$ \\
		F3 & -0.59 & -1.66 & 28.1 & $26.5 \rightarrow 29.8$ \\
		& & & & 32.2 \\
		\hline
	\end{tabular}
\end{table}

\subsection{Deprojection\label{Sect:Deproj}}

In order to model the different structures in the CO envelope of $\mu$~Cep, we deprojected the velocity cube produced from the NOEMA data. By assuming that the stellar wind has a constant velocity or is accelerating, it is possible to convert the (R.A., Dec., $v_\mathrm{LSR}$) coordinates into the ($x, y, z$) spatial coordinates relative to the star position. This is crucial to determine the structure sizes in space and ultimately their mass using radiative transfer modelling in Sect.~\ref{Sect:MassLoss}.

To perform the deprojection and obtain the actual sizes of the features, we first need to know the distance at which $\mu$~Cep is located. We saw in Sect.~\ref{Sect:Intro} that this parameter spans a large range in the literature. With a parallax estimate of $0.4778\pm 0.4677$~mas, the Gaia DR2 measurement \citep{2016A&A...595A...1G,2018A&A...616A...1G} cannot improve the distance estimate due to the large uncertainty. We will rely on the method used by \citet{2005A&A...436..317P}. Their value of $390 \pm 140$~pc is based on physical considerations on the relative sizes of the MOLsphere between the two M2I red supergiant stars $\mu$~Cep and Betelgeuse ($\alpha$ Ori). However, at that time, the distance of Betelgeuse was underestimated. \citet{2017AJ....154...11H} derived a distance of $222^{+48}_{-34}$~pc for $\alpha$ Ori. This enables us to scale the distance of $\mu$~Cep to $641^{+148}_{-144}$~pc.

Following the method of \citet{2018A&A...614A..12M}, we used the photometry from \citet{2002yCat.2237....0D}, the intrinsic extinction of a M2Ib star from \citet{1985ApJS...57...91E}, the interstellar extinction from \citet{1979ARA&A..17...73S}, estimated to be $A_\mathrm{V} = 2.08$ in the visible towards $\mu$~Cep, and the angular diameter derived from infrared interferometry from \citet{2005A&A...436..317P} to derive the stellar parameters (see Table~\ref{Tab:stellar_param}). The initial mass of the star was estimated from the stellar models of \citet{2012A&A...537A.146E}, and $\mu$~Cep is matching a model including rotation (Fig.~\ref{Fig:evol} in appendix).


\begin{table}
	\centering
	\caption{New estimation of the stellar parameters of $\mu$~Cep. The initial mass estimation comes from the comparison with evolutionary tracks from \citet{2012A&A...537A.146E} and plotted on Fig.~\ref{Fig:evol}.\label{Tab:stellar_param}}
	\begin{tabular}{ l l }
		\hline
		Stellar parameters & Values\\
		\hline\noalign{\smallskip}
		d & $641^{+148}_{-144}$~pc \\
		R$_\star$ & $972 \pm 228~\rso$ \\
		F$_\mathrm{\textit{UBVRIJHKLN}}$ & $(1.06 \pm 0.14) \tento{-8}$~W.m$^{-2}$\\
		T$_\mathrm{eff}$ & $3551 \pm 136$~K\\
		$\log $L$/\lso$ & $5.13^{+0.17}_{-0.28}$\\
		M$_\mathrm{init}$ & $15-20~\mso$ \\
		\hline
	\end{tabular}
\end{table}

Knowing the distance $d$ to $\mu$~Cep, the spatial coordinates ($x, y$) of an element can be determined from the right ascension and declination offsets ($\Delta \alpha, \Delta \delta$) :

\begin{equation}
	\begin{array}{l}
		x = d \times \Delta \alpha\\
		y = d \times \Delta \delta
	\end{array}
\end{equation}

\begin{figure}
	\includegraphics[width=\columnwidth]{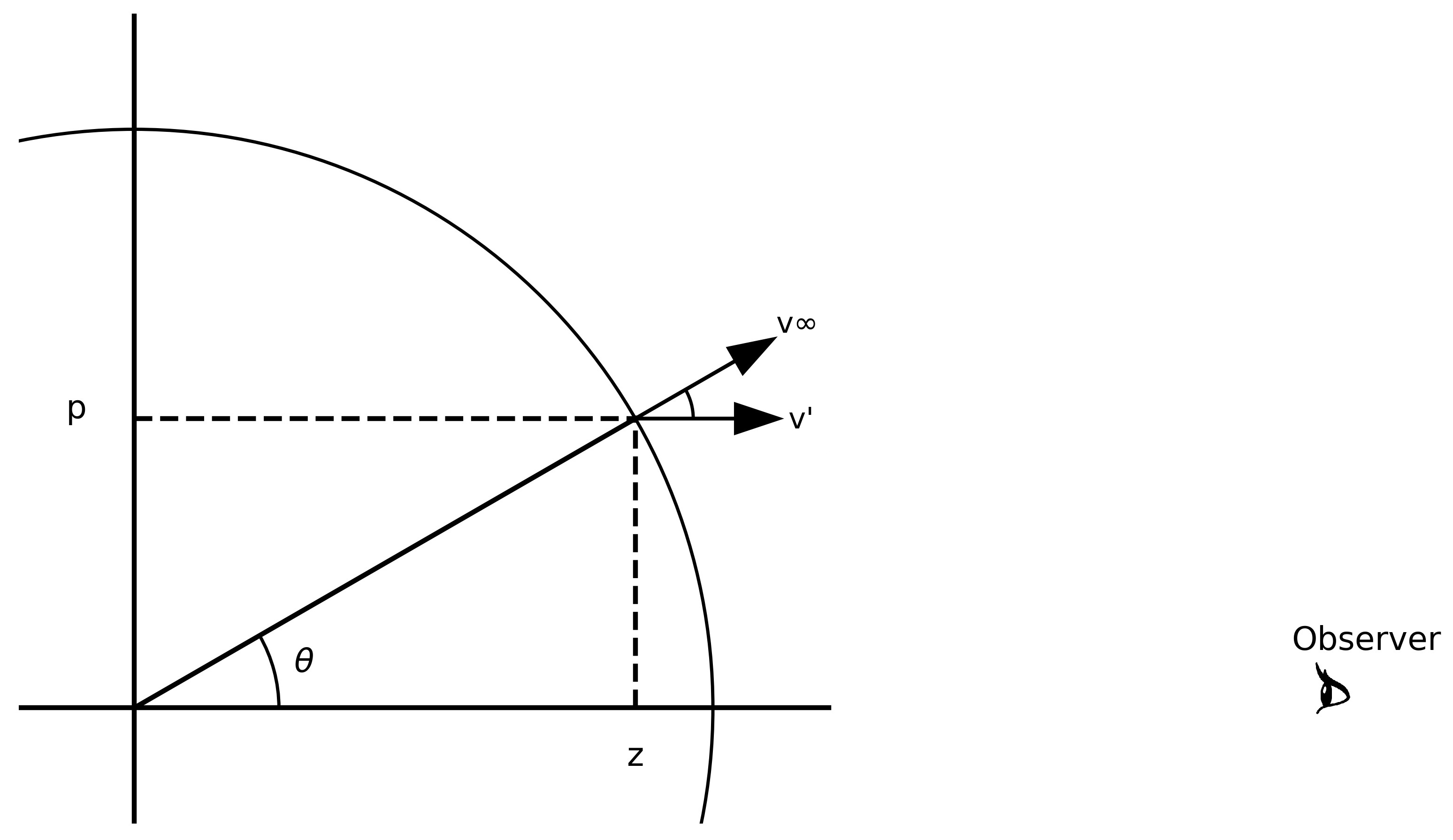}
	\caption{Principle of the deprojection from radial velocity space to full 3D space assuming a constant radial expansion velocity field of the material. The impact parameter $p$ corresponds to $\sqrt{x^2 + y^2}$.}
	\label{Fig:deproj_princip}
\end{figure}

To obtain the depth $z$ of a ($x, y, v_\mathrm{LSR}$) element, we use the same deprojection method as \citet{2018A&A...610A...4G}. In agreement with \citet{2016A&A...594A.108H}, we assume that the wind is expanding radially away from the star for distances greater than 5~$\rst$. From the modelling of Fe\,\textsc{ii} emission lines, \citet{2001ApJ...551.1073H} showed that the main acceleration of the wind of the prototypical M2I star Betelgeuse happened below 14~R$_\star$. For $\mu$ Cep this means inside the central 0.10~arcsec. Our beam  has a size of $0.92 \times 0.72$~arcsec or $590 \times 462$~au at 641~pc. Therefore we expect the acceleration zone to be only marginally resolved and we choose a constant velocity field. Based on simple geometric considerations (Fig.~\ref{Fig:deproj_princip}), we can derive:

\begin{equation}
	z = \sqrt{x^2+y^2} \frac{v'}{\sqrt{v_\infty^2 - v'^2}}
\end{equation}

%

\noindent with $v_\infty$ the terminal velocity, $v'= v - v_\star$ where $v_\star$ the systemic velocity of the star, and $v$ the radial velocity of the considered channel. \citet{2010A&A...523A..18D} derived a terminal wind velocity of $35~\kms$ by fitting the CO $J = 3-2$ and $J = 4-3$ line profiles. However, their estimation of the stellar wind velocity has been based on the assumption of a symmetric profile. With a LSR velocity of $32.7~\kms$ (App.~\ref{Sect:star_vel}), the CO $J = 2-1$ profile of $\mu$~Cep is clearly asymmetric (Fig~\ref{Fig:spectrum}). We define two terminal wind velocities: $v_\infty^\mathrm{slow} = 25.0~\kms$ determined by the red wing width, $v_\infty^\mathrm{fast} = 43~\kms$ determined by the blue wing width. We assume that the material within the velocity interval $v_\star \pm v_\infty^\mathrm{slow}$ belongs to a slow wind, and the rest is considered to be part of a faster wind directed toward the observer.

With these considerations, we can determine the full ($x, y, z$) coordinates of a light element in our cube. To obtain a proper representation, we transfer these $z$ values into a regular grid. The emission intensity from a cell of the original ($\Delta \alpha$, $\Delta \delta$, $V_\mathrm{LSR}$) cube is then put into the closest corresponding cell of the new ($\Delta x$, $\Delta y$, $\Delta z$) array. The result of this deprojection is shown in Fig.~\ref{Fig:3D_deproj}.

\begin{figure*}
	\includemedia[
	width=1\columnwidth,
	height=0.2\paperheight,
	addresource=deproj_3D.mp4,
	activate=onclick,
	deactivate=onclick,
	flashvars={%
	source=deproj_3D.mp4
	&scaleMode=stretch}
	]{\includegraphics[width=\columnwidth]{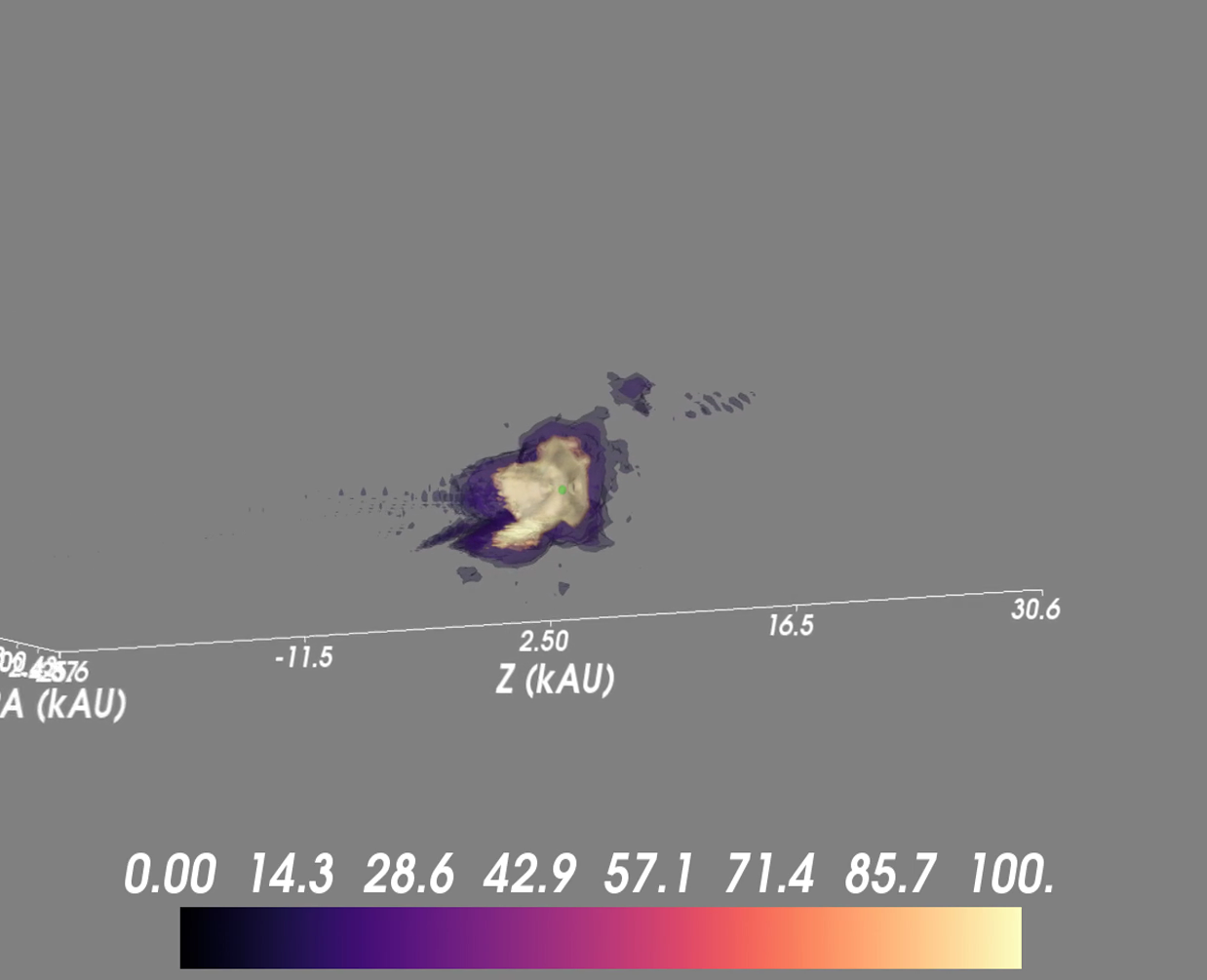}}{VPlayer.swf}
	\caption{3D rendering of the deprojection of $\mu$~Cep's environment in the CO $J=2-1$ line. An on-line version is available at: \url{https://frama.link/muCep_3D}.\label{Fig:3D_deproj}}
\end{figure*}

\section{Analysis of the mass loss}\label{Sect:MassLoss}

\subsection{Numerical methods}

To model the observed CO emission in the CSE of $\mu$~Cep we made use of the radiative transfer (RT) code {\sc lime}, which solves the equation of radiation transport in three dimensions under non-local thermodynamical equilibrium (NLTE) conditions. A comprehensive description of the schemes and techniques used in the code can be found in \citet{2010A&A...523A..25B}. The model is sampled by $5\tento{4}$ grid points; half of which are distributed logarithmically (with a gradual grid refinement towards the centre), and half are distributed randomly. Of the latter, the position of the grid points is weighted by relative density, further increasing the mesh refinement in the high-density regions of the model. An additional $5\tento{3}$ grid points define the edge of the numerical domain. In these points the RT is concluded by assuming the addition of the microwave background intensity. These $5.5\tento{4}$ grid points are Delaunay-triangulated, and subsequently Voronoi-tesselated in order to create the cells around each grid point, inside which the physical conditions are assumed to be constant (determined by the defining grid point). The physical set-up is described in terms of density, temperature, molecular abundance, macro- and micro-scale velocity fields. 

The mean intensity field is determined by solving the RT equations along the Delaunay lines connecting the neighbouring grid points, extending to the edge of the numerical domain. This mean intensity field serves to update the level populations in each grid point. When these have converged, a ray tracing algorithm produces frequency dependent intensity maps, as viewed from a user-defined vantage point. We note that the user-defined physical set-up (density, temperature, abundances, etc) is not adapted as the level populations converge.

{\sc lime} only allows for one dust species to be included in the calculation, for which only thermal emission is calculated. Hence, the code is only suitable for rudimentary first order dust modelling. Furthermore, because the CO molecule has such a low electric dipole moment, the contributions of the surrounding diffuse radiation field (from e.g. the dust) to the level excitation are not very important. This has been checked by assuming a distribution of amorphous silicate dust that follows the gas density (assuming a gas-to-dust mass-ratio of 200), which has been found to have no effect on the final emission distribution. This finding reduces the available free parameters and thus simplifies the modelling substantially. We have therefore opted to omit any dust contributions from the RT model. The spectroscopic CO data of the LAMDA database \citep{2005A&A...432..369S} were used (with 82 levels in $v=0$, and no levels in $v>0$ ); the collisional rates were taken from \citet{2010ApJ...718.1062Y}.

After having retrieved the intrinsic emission distribution from {\sc lime}, we have post-processed it with {\sc casa} \citep{2007ASPC..376..127M} in order to simulate a synthetic NOEMA observation. The actual general observation conditions and instrumental set-up have been adopted as input parameters for the simulations, and are given in Table \ref{casa}. In these simulations we have used the same antenna locations and diameters as for the observation to ensure consistent ($u, v$) plane coverage.

\begin{table}
	\centering
	\caption{The NOEMA observation simulation specifications. \label{casa}}
	\begin{tabular}{ l l }
		\hline
		\multicolumn{2}{ c }{Simulation Parameters} \\
		\hline
		Pixel size of input model & 0.077~arcsec \\
		Field size of input model & 15.5~arcsec \\
		Peak flux & Taken from {\sc lime}  \\ 
		Thermal noise & Standard \\
		Write-out time for single visibility point & 45 sec \\
		Integration time & 201 min on-source \\
		\hline
	\end{tabular}
\end{table}

\subsection{Modelling strategy\label{Sect:model_stra}}

To reproduce the emission features presented in the channel maps (Fig.~\ref{Fig:channel_maps}) we require a description of the four primary attributes of the stellar wind in each point of the CSE: the velocity, the molecular abundance, the temperature, and the density. However, because $\mu$~Cep has not been studied in great detail in the past, tight constraints on the above-mentioned wind properties are non-existent in the literature. Yet, combining the literature with the analysis of the current data we successfully constructed a CSE model, which is described below.

\subsubsection{Velocity}

Following the reasoning outlined in Sect.~\ref{Sect:Deproj}, we assume the velocity of the wind to be defined as constant, with a velocity of 25~$\kms$, throughout the entire CSE, except for the region in which the clump C1 is located. In this region we have assumed that the velocity is 43~$\kms$. We elaborate on the potential consequences and limits of these assumptions in Sect.~\ref{Sect:Disc_Vel}.

\subsubsection{Molecular abundance}

Not much is known about the typical molecular abundances in the CSEs of red supergiant stars. The CSE of $\mu$ Cep was studied by \citet{2010A&A...523A..18D}, as part of a larger sample for which, through the exploration of a parameter grid with one-dimensional radiative transfer models, a number of critical CSE properties were derived. They constrained the fractional CO abundance with respect to H to be between 1 and 5 $\tento{-4}$, being the limits of their parameter grid. For this exercise we adopt the CO/H$_2$ abundance to be 1.5$\tento{-4}$.

Betelgeuse has the same spectral type as $\mu$ Cep, thus we reasonably assume that its circumstellar chemistry is similar to that of $\mu$ Cep. Hence, if we consider all carbon to be locked up into CO, we find our assumed value to be in agreement with the atomic carbon estimates in the photosphere of Betelgeuse \citep[$2.5 \tento{-4}$,][]{1984ApJ...284..223L,1986ApJ...306..605G}.

However, because red supergiant stars tend to have chromospheres with temperatures high enough to destroy molecules, there is a possibility that the currently assumed CO abundance is severely overestimated. We address this issue and its implications to the models in Sect. \ref{Sect:CO_Abund}.

\subsubsection{Temperature:}

We have not been successful in finding a detailed description of the temperature profile throughout the wind of $\mu$~Cep. Hence, because of the lacking understanding of the nature of supergiant winds, we have approximated the temperature profile in the stellar wind as being dominated by adiabatic cooling, assuming a simplified spherically symmetric model. An adiabatic process is characterised by the invariance of the quantity $TV^{\gamma - 1}$, where $T$ is the temperature of the gas, $V$ is its volume, and $\gamma$ is the adiabatic index of the gas, which takes the value of $5/3$ for a monoatomic gas. Thus, for a freely expanding spherical shell of gas we recover that the gas temperature $T(r)$ is described as
\begin{equation}
	 T(r) = T_* \left(\frac{r}{R_*}\right)^{-1.33},
\end{equation}
where $r=\sqrt{x^2+y^2+z^2}$ is the radial coordinate, $T_*$ is the stellar surface temperature, and $R_*$ is its diameter.

One might argue that it would be more reasonable to adopt the temperature value of a well studied red supergiant like Betelgeuse, which has been caulculated by \citet{1991ApJ...382..606R} taking into account a range of different heating and cooling components. However, \cite{2017ApJ...836...22H} showed that this model has some biases. In particular, it overestimates the temperature at its inner boundary (3~R$_\star$). Moreover, from SOFIA-EXES observations, \citet{2018AAS...23232501H} demonstrated that the chromospheric activity of $\mu$~Cep was weaker than Betelgeuse's. Both remarks drag the temperature profile of \citet{1991ApJ...382..606R} towards an overestimation of $\mu$~Cep's. Finally, the differences between this complex profile and the simple adiabatic one are stronger close to the star, in the central beam area of our NOEMA maps. This region was modelled as a smooth outflow (Sect. \ref{Sect:density}), this is the simplest hypothesis, however it is most likely a conglomerate of unresolved clumps (Sect \ref{Sect:RT_results} and \ref{Sect:Disc_density}). For these reasons, in order to avoid overconstraining our model, which could create more non-verifiable biases, we decided to use the simplest adiabatic temperature profile.

\subsubsection{Density\label{Sect:density}}

To model the density distribution in the channel maps, the wind density is partitioned into two specific contributions. The first contribution comes from a sequence of three-dimensional Gaussian clumps with a density $\rho_{\rm i}$ which is given by 
\begin{equation}
	 \rho_{\rm i} = \rho_{\rm i, max}{\rm exp}\left[-\frac{(x-x_i)^2}{2s_{x,i}^2}-\frac{(y-y_i)^2}{2s_{y,i}^2}-\frac{(z-z_i)^2}{2s_{z,i}^2}\right],
	 \label{Eq:model_clumps}
\end{equation}
where $x,y,z$ are the Cartesian coordinates (with the origin at the stellar location) with $x$ pointing along the positive RA, $y$ along the positive DEC, and $z$ along the line of sight (positive when moving away from the observer); $x_i,y_i,z_i$ is the location of the centre of clump $i$ where it has a density $\rho_{\rm i, max}$; and $s_{x,i},s_{y,i},s_{z,i}$ is its one sigma Gaussian width in each dimension.

We attempt to reproduce the most prominent clump emission by measuring their position as observed in the CSE of $\mu$~Cep, summarized in Table~\ref{Tab:Feature_Id}. The first guess for the position is measured in the deprojected emission cube (Fig~\ref{Fig:3D_deproj}), and further adjusted in velocity-space to match the observations. In most cases, the sizes are found to be below the angular resolution provided by the present NOEMA observations. Therefore, they are first set to the beam size and left as a free parameter to match the observations.

The second component of the emission emerges from what we observe to be an unresolved central emission. We model this as a smooth outflow, for which the density $\rho_{\rm so}$ is given by 
\begin{equation}
	 \rho_{\rm so} = \frac{\dot{M}_{\rm so}}{4 \pi r^2 v_w(r)},
\end{equation}
where $r=\sqrt{x^2+y^2+z^2}$ is the radial coordinate, $\dot{M}_{\rm so}$ is the mass loss rate through the smooth outflow, $v_w(r)$ is the velocity of the slow wind taken as $25~\kms$ (see Sect.~\ref{Sect:Deproj} and further discussion in Sect.~\ref{Sect:RT_results}). 

\subsection{Radiative transfer results\label{Sect:RT_results}}

Under the above-mentioned assumptions we reproduce the emission of the most prominent clumps in terms of both maximum intensity and integrated flux density. The best matching clump sizes, densities and masses are summarised in Table~\ref{RT}. We show their synthetic emission distribution in Fig.~ \ref{Fig:clump_spec}. As stated in Sect.~\ref{Sect:density}, most of the clumps are unresolved but their size within the model was left as a free parameter with the beam size as initial guess. The final values in Table~\ref{RT} correspond to best matching sizes and we can only appreciate that they are unresolved in the ($x, y$) directions.

\begin{figure*}
	\centering
	\includegraphics[width=\textwidth]{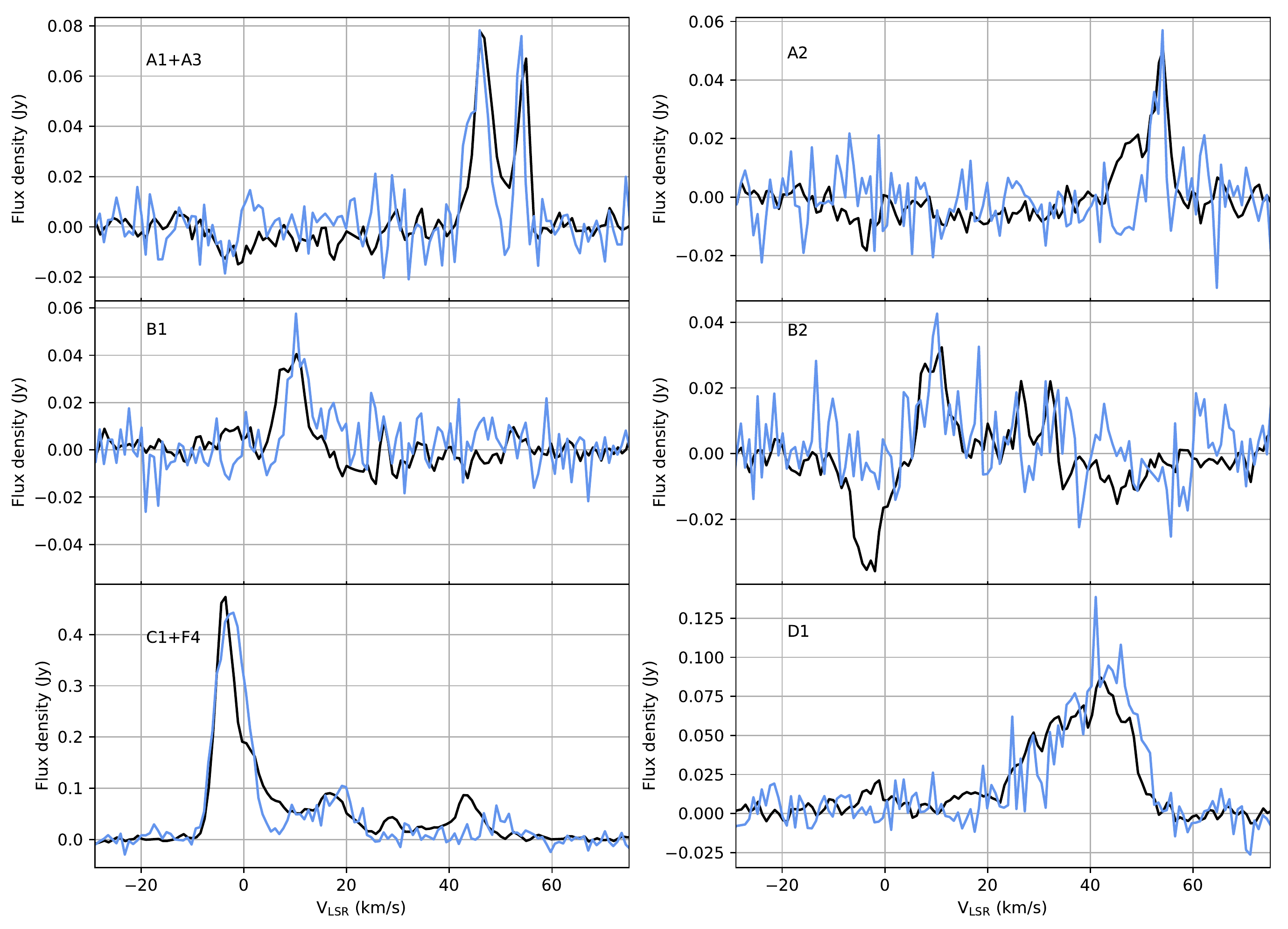}
	\caption{Synthetic emission of the clumps in light blue compared to the NOEMA observations in black. See Table~\ref{RT} for their detailed modelling. \label{Fig:clump_spec}}
\end{figure*}

As there is no reason to  exclude that the central unresolved area of the channel maps (Fig.~\ref{Fig:channel_maps}) is hosting other clumps, it is not easy to accurately model it as a smooth outflow component. However, if we make the assumption that the high-velocity emission around the position of the stellar source (Fig.~\ref{Fig:spectrum}, red curve) originates from a smooth outflow material not contaminated by clumps, then it is possible to obtain an estimate of its mass-loss rate. From the channel maps (Fig~\ref{Fig:channel_maps}), it appears that the central source is more asymmetric in the blue channels compared to the red ones. We can expect that the blue channels carry the greatest part of the clump contamination. Therefore, we attempt to reproduce only the high-velocity red-shifted secondary horn (centred at $v_\mathrm{LSR} \sim 45~\kms$, Fig. \ref{Fig:spectrum}, red curve) of the expected double-peaked component of the spectrum with a smooth outflow. By doing this, we obtain an estimate for the mass-loss rate through the unresolved central component of $2\tento{-6}~\msoy$. From a comparison with the spectrum shown in \citet{1989A&A...210..198L}, we estimate that no large scale structure was filtered out by the interferometer. Based on the reproduction of the red horn of the central spectrum, we estimate the uncertainty on the derived smooth outflow mass-loss rate to be dominated by the flux calibration uncertainty ($\le 20\%$).

\begin{table*}
	\centering
	\caption{Results of the radiative transfer modelling with \textsc{lime} for the different clumps. See Table~\ref{Tab:Feature_Id} and Fig.~\ref{Fig:channel_maps} for a correspondence of the clump ids. Positions and sizes of the clumps are defined in Eq.~\ref{Eq:model_clumps}.\label{RT}}
	\sisetup{
		table-number-alignment = center,
		table-figures-integer  = 2
	}
	\begin{tabular}{c S S S S S S S[table-figures-exponent = 1] S}
		\hline
		Id. & {$x_i$} & {$y_i$} & {$z_i$} & {$s_{x,i}$} & {$s_{y,i}$} & {$s_{z,i}$} & {$\rho_{i,\mathrm{max}}$} & {Mass} \\
		$i$ & {(10$^{\rm 3}$AU)} & {(10$^{\rm 3}$AU)} & {(10$^{\rm 3}$AU)} & {(10$^{\rm 3}$AU)} & {(10$^{\rm 3}$AU)} & {(10$^{\rm 3}$AU)} & {(kg\,m$^{-3}$)} & {(10$^{\rm -5}\mso$)} \\
		\hline \noalign{\smallskip}      
		A1 & -0.95 & 2.20 & 4.30 & 0.22 & 0.17 & 0.27 & 4.5e-17 & 2.0 \\
        
		A2 & -1.50 & 1.90 & 4.00 & 0.22 & 0.19 & 0.48 & 3.5e-17 & 2.8 \\
			
		A3 & -0.75 & 2.10 & 1.50 & 0.17 & 0.13 & 0.30 & 4.0e-17 & 1.2 \\
        
		B1 & 1.90 & -0.40 & -4.50 & 0.29 & 0.26 & 2.40 & 1.0e-17 & 8.1  \\
        
		B2 & 1.30 & -1.05 & -3.60 & 0.22 & 0.19 & 2.40 & 1.0e-17 & 4.3  \\
        
		C1 & -0.55 & -1.05 & -1.40 & 0.34 & 0.31 & 0.15 & 2.5e-16 & 17.2  \\
        
		D1 & -0.25 & 1.27 & 0.35 &  0.34 & 0.28 & 0.48 & 4.0e-17 & 8.1  \\
        
		F4 & -0.35 & -0.60 & -0.50 &  0.15 & 0.21 & 0.09 & 1.5e-16 & 1.8  \\
        
		\hline
	\end{tabular}
\end{table*}

\section{Discussion}\label{Sect:Discussion}

\subsection{Density distribution throughout the outflow \label{Sect:Disc_density}}

In Sect. \ref{Sect:RT_results} we model most of the red-shifted emission in the central aperture spectrum by assuming a homogeneous smooth outflow (HO) with mass-loss rate of $2 \tento{-6}~\msoy$. However, looking at the more blue-shifted channels in our model (red curve of Fig.~\ref{Fig:spec_model}), the second horn of the double-peaked profile prominently peaks at $\sim 10~\kms$, whereas the data shows no strong emission around this velocity (red curve of Fig~\ref{Fig:spectrum}). This discrepancy can be explained by two scenarios. 

\begin{figure}
	\includegraphics[width=\columnwidth]{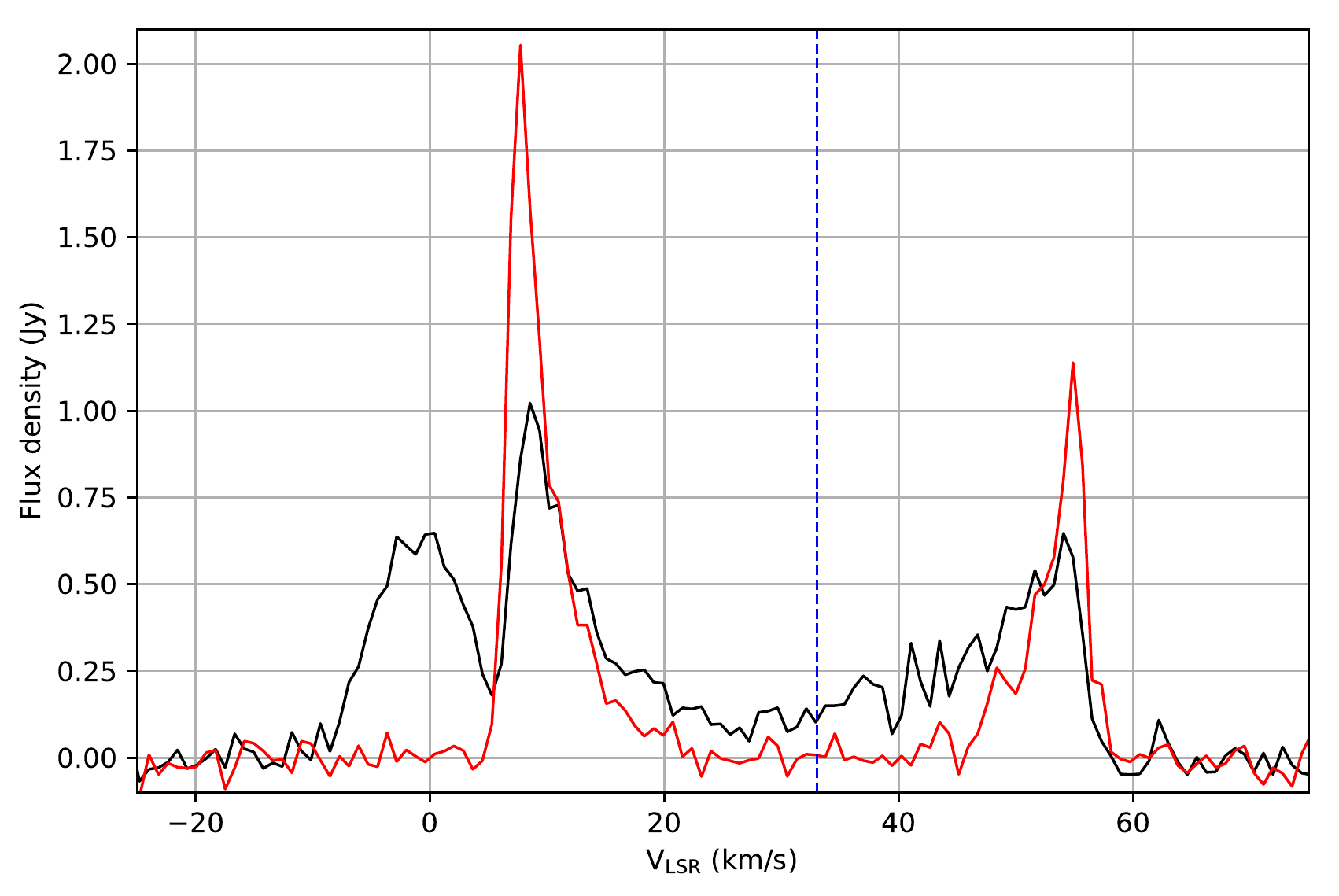}
	\caption{Integrated model line profile over the central $12\times12$~arcsec region of the continuum subtracted CO $J=2-1$ (230.538~GHz) maps of $\mu$~Cep in black. The red curve correspond to 10 times the model flux density integrated over a beam size aperture at the phase centre. The vertical dashed blue line corresponds to the star velocity derived in App.~\ref{Sect:star_vel}.\label{Fig:spec_model}}
	\centering
\end{figure}

Either this is a consequence of the fact that the phenomenon producing the clumps is affecting the blue-shifted mass loss, causing less gas to being ejected towards the observer.

Or, this could be an indication that there is simply no homogeneous component to the outflow, and that all emission originates from a conglomerate of differently sized clumps of gas. In fact, if each clumpy feature were modelled separately, and provided we had a proper description of the velocity field around the star, we would be able to successfully reproduce the full CSE emission distribution by simply assuming a carefully placed distribution of clumps. But since the inner clumps remain unresolved in the data, they cannot be properly modelled.

The nature of the central unresolved emission could only be determined with a better angular resolution, for example with the future capabilities of NOEMA in its most extended configuration.

\subsection{Mass loss}

\subsubsection{Velocity field of the gas\label{Sect:Disc_Vel}}

We are unable to properly model the emission distribution in clump C2 under the assumptions listed in Sect.~\ref{Sect:model_stra}. In particular, the assumption of a Gaussian density distribution along the line of sight completely fails to reproduce the spatial emission distribution of clump C2. 
This may either indicate that shape of clump C2 is of such nature that it cannot be approximated by a 3D Gaussian, or that clump C2 resides in the region where our assumptions on the local velocity field breaks down. Seeing as the assumption of a Gaussian density distribution does not seem unreasonable to first order (see Fig.~\ref{Fig:chan_map_model} in the appendix) for all other clumps, the latter scenario may indeed be the more likely one.

\subsubsection{Mass-loss rate estimate\label{Sect:Mass_loss_rate}}

We calculate the absolute lower limit on the mass-loss rate through the clumps by assuming that the small, unresolved, clumps do not contribute significantly to the total mass loss. The clumps modelled in Table \ref{RT} represent a total mass of $45.5 \tento{-5}~\mso$. Over a distance of $4.92 \tento{3}$~AU, and considering the velocity of $v^\mathrm{slow}_\infty = 25~\kms$ of the farthest clumps (A1 and A2), we arrive at lower mass-loss rate estimate through the clumps of $4.9 \tento{-7}~\msoy$. From the flux calibration of the observations, we apply a 20\% uncertainty of $\pm 1.0 \tento{-7}~\msoy$, yielding a lower limit for the total mass-loss rate including the smooth outflow of $(2.5 \pm 0.5) \tento{-6}~\msoy$.

Concerning the clumps that were not modelled, the only assumption we could make is estimating their mass to be the mean mass of the modelled clumps ([$5.69 \pm 5.38] \tento{-5}~\mso$). Doing so, we estimate the mass-loss through the clumps to be $(11.6 \pm 11.0) \tento{-7}~\msoy$. This value has a poor meaning due to its large uncertainty and the extrapolation performed on the smaller clumps.

\citet{2010A&A...523A..18D} estimate a mass-loss rate of $2~\tento{-6}~\msoy$. This is based on the previous distance estimation, and on spatially unresolved (3-2) and (4-3) CO line profiles that have been modelled assuming a smooth spherical outflow. In the context of embedded spiral morphologies \citet{2015A&A...579A.118H} have shown that the misinterpretation of a three-dimensional density distribution as an effectively one-dimensional outflow can yield errors on the derived mass-loss rates of up to a factor 10. This argument can be generalized to other morphologies. Although within the uncertainties of both the model presented here and the limitations of the modelling by \citet{2010A&A...523A..18D}, the results are effectively compatible, we will not compare them any further due to the very different underlying assumptions.

\subsubsection{CO abundance and mass loss rate\label{Sect:CO_Abund}} 

Our assumed fractional CO abundance value may be severely overestimated due to the presence of the chromosphere around $\mu$~Cep. \citet{1994ApJ...424L.127H} argue that in such cases most of the carbon in the CSE is found in atomic form, resulting in typical CO abundances that are a factor of 10 lower. This will strongly impact the derived clump mass values, and hence also the newly deduced mass-loss rate. However, all modelled clumps except for C1 have been found to be optically thin, which implies that their masses will scale inversely with the factor rescaling the assumed CO/H$_{\rm 2}$ fraction. Hence, assuming a factor ten decrease in molecular abundance will result in an increase of the deduced clump mass by a factor ten. We cannot extend this argument to the mass contained by C1. However, if we assume that a similar process has indeed created all clumps (including C1), and that this process tends to eject parcels of matter of similar mass, then we could assume that the mass in clump C1 would also scale inversely with the assumed CO/H$_{\rm 2}$ ratio. Consequently, for a CO abundance lower by a factor 10, the mass-loss rate via the clumps would be as high as $(4.9 \pm 1.0) \tento{-6}~\msoy$, a value comparable to the total mass-loss rate derived by \citet{2016AJ....151...51S}.

\subsubsection{Mass-loss mechanism(s)\label{Sect:MassLoss_Mechanism}}

In Sect.~\ref{Sect:Mass_loss_rate} we calculated that the lower estimate of the mass-loss rate through the clumps is $(4.9 \pm 1.0) \tento{-7}~\msoy$. The clump masses range between 1.2 and 17.3 $\tento{-5}~\mso$.

Dynamical models of evolved star atmospheres show that the large and smaller-scale convective cells that circulate in the stellar mantle continuously break the stellar surface, rendering it highly dynamical on a range of different time scales \citep{2011A&A...535A..22C}. If the thermal, dynamical, and potentially magnetic, processes that generate these large-scale motions conspire to occasionally produce a surface burst of enhanced potency, then it is possible that it results in the highly directed ejection of stellar matter \citep{2018A&A...609A..67K}. Such mass-loss process could be the origin of the observed clumps. However they cannot account for the smooth outflow component, so it appears that there must be at least another mechanism extracting material from the star, except if the smooth wind is itself made of smaller clumps that remain unresolved with the current observations.



\subsection{Comparison with other nearby red supergiant stars}

With its mass loss of $(2.5 \pm 0.5) ~\tento{-6}~\msoy$, $\mu$~Cep  is one of the RSGs with a tenuous outflow, like Betelgeuse ($\alpha$~Ori, M2Ia-Iab, $1.0 \tento{-6}~\msoy$ according to \citealt{2011A&A...526A.156M}, $2.2 \tento{-6}~\msoy$ assuming the star at 200~pc for \citealt{1987ApJ...315..305B}, or $4 \tento{-6}~\msoy$ according to \citealt{1990A&A...227..141M}) and unlike VY~CMa for example (M5Iae, $3 \tento{-4}~\msoy$, \citealt{2005AJ....129..492H}).

The mass-loss history of VY CMa has been previously investigated through spectroscopy and radiative transfer modelling by \citet{2006A&A...456..549D}. It appears that the star went through several mass-loss episodes with different mean mass loss rates: from $\sim 1 \tento{-6}~\msoy$ to $\sim 3.2 \tento{-4}~\msoy$. These episodes may have lasted from 100~yr for the most intense to 800~yr for the lowest mass loss rates. The circumstellar environment of VY CMa has been observed in various spectral domains and always appears inhomogeneous and asymmetric: with HST/WFPC2 \citep{2007AJ....133.2716H}, VLT/SPHERE \citep{2015A&A...584L..10S} and also with ALMA \citep{2015A&A...573L...1O}. These last authors conclude that the dust clumps they observed were emitted through a period of 30 to 50~yr. They conclude that they cannot have been emitted by photospheric convection as these cells are expected to have a lifetime of several months at best \citep{2011A&A...535A..22C}. It should be noted that these features are not similar to the gaseous structures we observed in the CSE of $\mu$~Cep, in particular they are much more massive. 

Betelgeuse and $\mu$~Cep share a common spectral type. The CSE of Betelgeuse was imaged in the optical by \cite{2009A&A...504..115K,2011A&A...531A.117K}. Various structures have been revealed and point towards an episodic and inhomogeneous mass loss. Clumps have been observed in the optical and infrared, for example in the K I line \citep{2002A&A...386.1009P}, Na I and K I \citep{1995A&A...298..869M}, and CO at 4.6~$\mu$m \citep{2009AJ....137.3558S}. Recently, \citet{2018A&A...609A..67K} have hypothesized that the poles of the star may be the location of long lived giant convective cells observed by \citet{2016A&A...588A.130M} and \citet{2017A&A...602L..10O} that could be responsible for the dusty structure observed in visible linear polarization with VLT/SPHERE \citep{2016A&A...585A..28K}. However, there is no theoretical model available to explain these observations. The CO $J=2-1$ transition in the CSE of Betelgeuse was observed with CARMA \citep{2012AJ....144...36O}. Their restoring beam and spectral resolution were similar to our NOEMA observations of $\mu$~Cep. The similarity of the clumpiness of the circumstellar environments is remarkable. It is a possible indication that the same processes are at play to trigger the mass loss of both stars and possibly of all RSG experiencing this low mass-loss regime.


\section{Conclusion}\label{Sect:Conclusion}

Through NOEMA CO $J=2-1$ observations, we obtained a detailed overview of the gaseous envelope of the RSG star $\mu$~Cep. Its circumstellar environment contains several clumps. The channel maps were deprojected assuming two constant wind velocities in two different regions of the line of sight. The deprojected cube is a full 3D representation of the CO environment of $\mu$~Cep. We modelled this outflow using the radiative transfer code \textsc{lime}. This allows to estimate the mass-loss rate of the star to be $(2.5 \pm 0.5) \tento{-6}~\msoy$, of which $(4.9 \pm 1.0) \tento{-7}~\msoy$ are due to the clumps. Moreover, assuming a plausible CO abundance lower by a factor 10 (that could be caused by the chromosphere), the mass loss rate through the clumps could be ten times more important. Therefore, the clump contribution to the mass loss is quite significant ($\ge 25~\%$). This modelling questions the nature of any smooth outflow component as most of the clumpy environment can be reproduced with simple Gaussian symmetric structures, except for the central features where we are limited by the degeneracy inherent to the deprojection method and the resolving power of the interferometer.

The very inhomogeneous distribution of the clumps and their localized nature is a strong argument in favour of episodic mass loss events around $\mu$~Cep. We note that this star behaves in various spectral domains like the prototypical M2I RSG Betelgeuse. The current best scenario to explain these observations would be a convection triggered mass loss but several observations \citep{2015A&A...575A..50A,2015A&A...573L...1O} cannot be reproduced by current convective models. However, only a small sample of red supergiant convective models is currently available: more simulations based on various recipes would be required for a proper conclusion on the role of convection.



\section*{Acknowledgements}

We are greateful to the referee whose detailed comments improved the quality of this paper. 
We would like to thank Dr.~A.~Avison from the UK ALMA node with his help with \textsc{casa}. 
This work has benefited from important contributions from Dr. A. Castro-Carrizo and Dr. V. Bujarrabal. 
This work is based on observations carried out under project number W15BL with the IRAM NOEMA Interferometer. IRAM is supported by INSU/CNRS (France), MPG (Germany) and IGN (Spain). We are grateful to the IRAM/NOEMA staff at the Plateau de Bure observatory for the successful execution of the observations. 
This project has received funding from the European Union's Horizon 2020 research and innovation program under the Marie Sk\l{}odowska-Curie Grant agreement No. 665501 with the research Foundation Flanders (FWO) ([PEGASUS]$^2$ Marie Curie fellowship 12U2717N awarded to M.M.). 
LD, WH, DK, and NC acknowledge support from the ERC consolidator grant 646758 AEROSOL. 
SS thanks the Belgian Science Policy Office for their support through contract no. BR/143/A2/STARLAB. 
GMH acknowledges infrastructure support from CU-CASA. 
We used the SIMBAD and VIZIER databases at the CDS, Strasbourg (France)\footnote{Available at \url{http://cdsweb.u-strasbg.fr/}}, and NASA's Astrophysics Data System Bibliographic Services. 
This research made use of Matplotlib \citep{Hunter:2007}, Astropy\footnote{Available at \url{http://www.astropy.org/}}, a community-developed core Python package for Astronomy \citep{2013A&A...558A..33A}, and  Uncertainties\footnote{Available at \url{http://pythonhosted.org/uncertainties/}}: a Python package for calculations with uncertainties.




\bibliographystyle{mnras}
\bibliography{biblio} 



\appendix

\section{Evolutionary tracks for $\mu$ Cep}\label{Sect:evol}

Figure~\ref{Fig:evol} shows the evolutionary tracks from \citet{2012A&A...537A.146E} in a Hertzsprung-Russell diagram, with the derived position of $\mu$~Cep from the new distance estimate of Sect.~\ref{Sect:Deproj}, Table~\ref{Tab:stellar_param}.

\begin{figure}
	\includegraphics[width=\columnwidth]{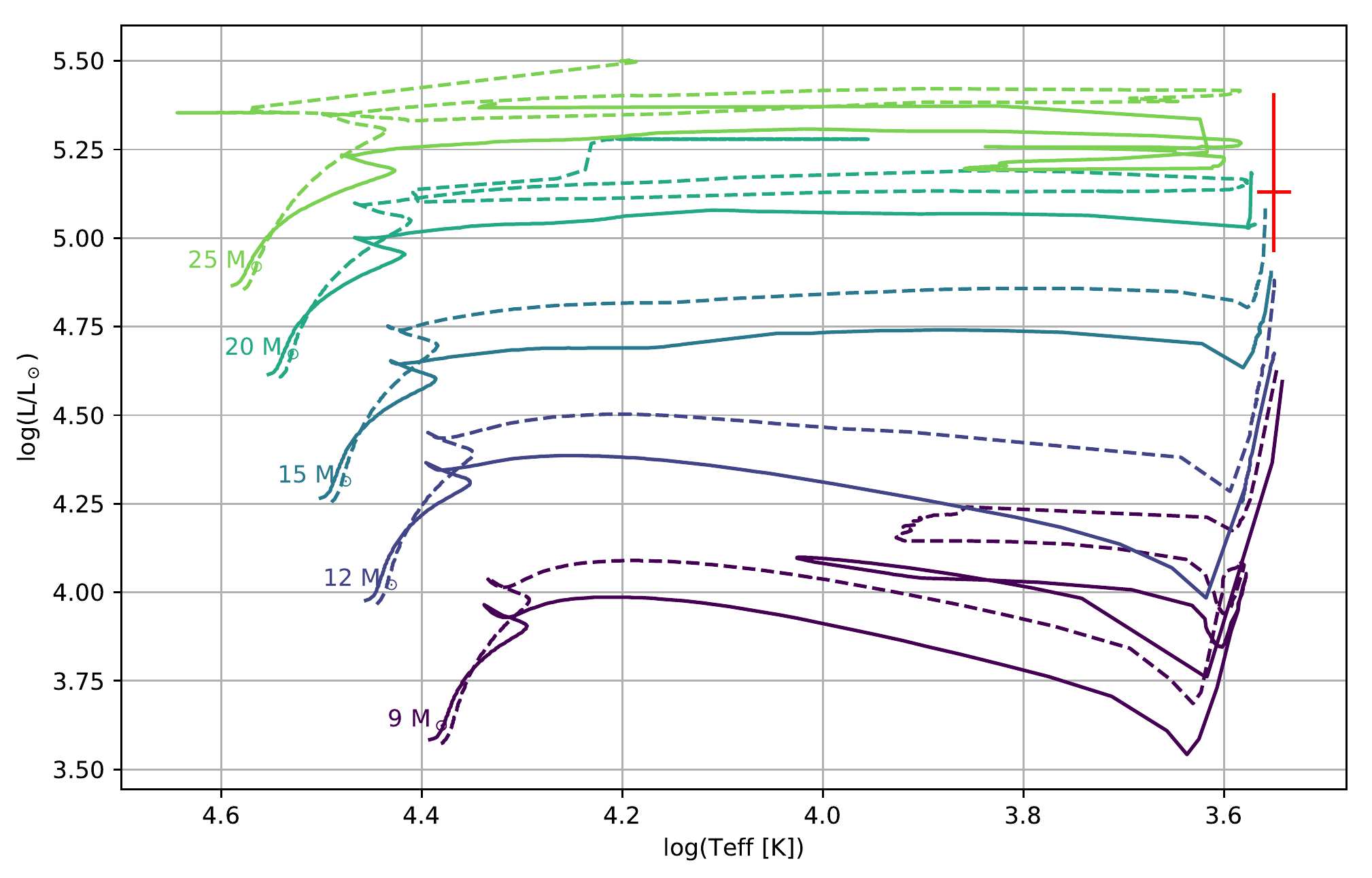}
	\caption{Hertzsprung-Russell diagram showing the evolutionary tracks from the model of \citet{2012A&A...537A.146E}. The continuous lines correspond to non-rotating models, and the dashed lines to models including rotation. The red cross corresponds to the newly derived stellar parameters of $\mu$~Cep derived in Sect.~\ref{Sect:Deproj} (Table \ref{Tab:stellar_param}).\label{Fig:evol}}
	\centering
\end{figure}

\section{Velocity of $\mu$ Cep }\label{Sect:star_vel}

We obtained a high SNR HERMES spectrum of $\mu$~Cep taken on the 6th of October, 2018 \citep{2011A&A...526A..69R}. The HERMES spectrograph is mounted on the 1.2~m Mercator Telescope at the Roque de los Muchachos Observatory, La Palma. HERMES spectra have a wavelength coverage of 380-900 nm with a spectral resolution of $R = 85000$. The radial velocity of $\mu$~Cep was obtained by cross-correlating the reduced HERMES spectra with a predefined software mask adapted for a given object type. We made use of a F0 mask that contains around 1200 lines obtained in spectral orders between 54 and 74 (477 - 655~nm). By fitting a Gaussian fit to the cross-correlation function and computing the mean of the fit, we obtained a radial velocity of $32.7~\kms$ for $\mu$~Cep in the LSRK frame. An error of $\pm$ $0.1~\kms$ which is one standard deviation of the Gaussian fit can be associated to this radial velocity.

\section{Results of the clumpy radiative transfer model\label{Sect:model_clump}}

The details and the analysis of the radiative transfer modelling are given in Sect.~\ref{Sect:MassLoss}. Fig.~\ref{Fig:chan_map_model} represents the corresponding channel maps. They were generated by simulating the observation of the model by the NOEMA interferometer using the same ($u, v$) sampling as the actual observations.

\begin{figure*}
	\includegraphics[width=\textwidth]{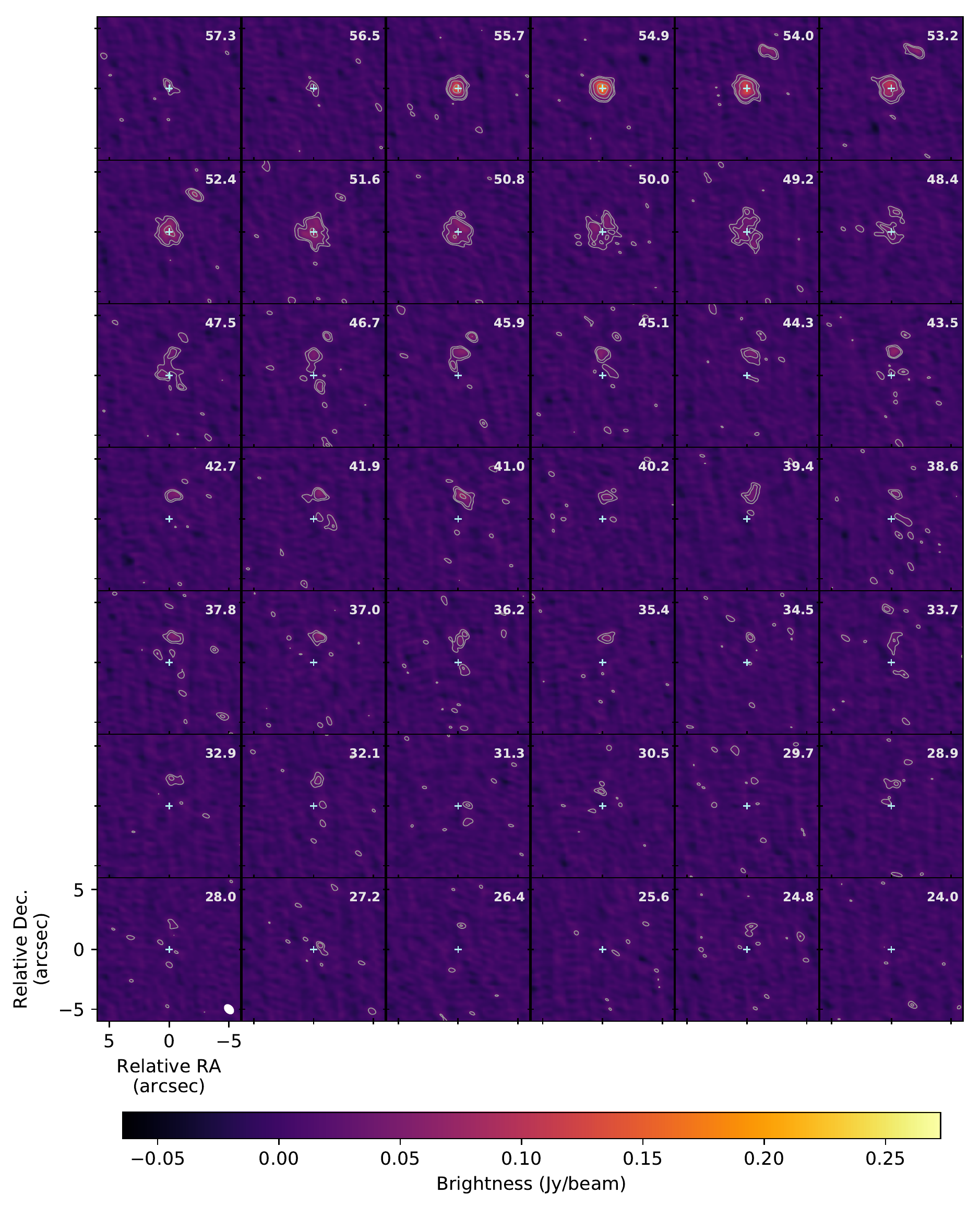}
	\caption{Cleaned channel maps of the clumpy CO environment of $\mu$~Cep resulting from the radiative transfer modelling (Sect.~\ref{Sect:MassLoss}). The plotting parameters are similar to Fig.~\ref{Fig:channel_maps}.\label{Fig:chan_map_model}}
	\centering
\end{figure*}

\begin{figure*}
	\includegraphics[width=\textwidth]{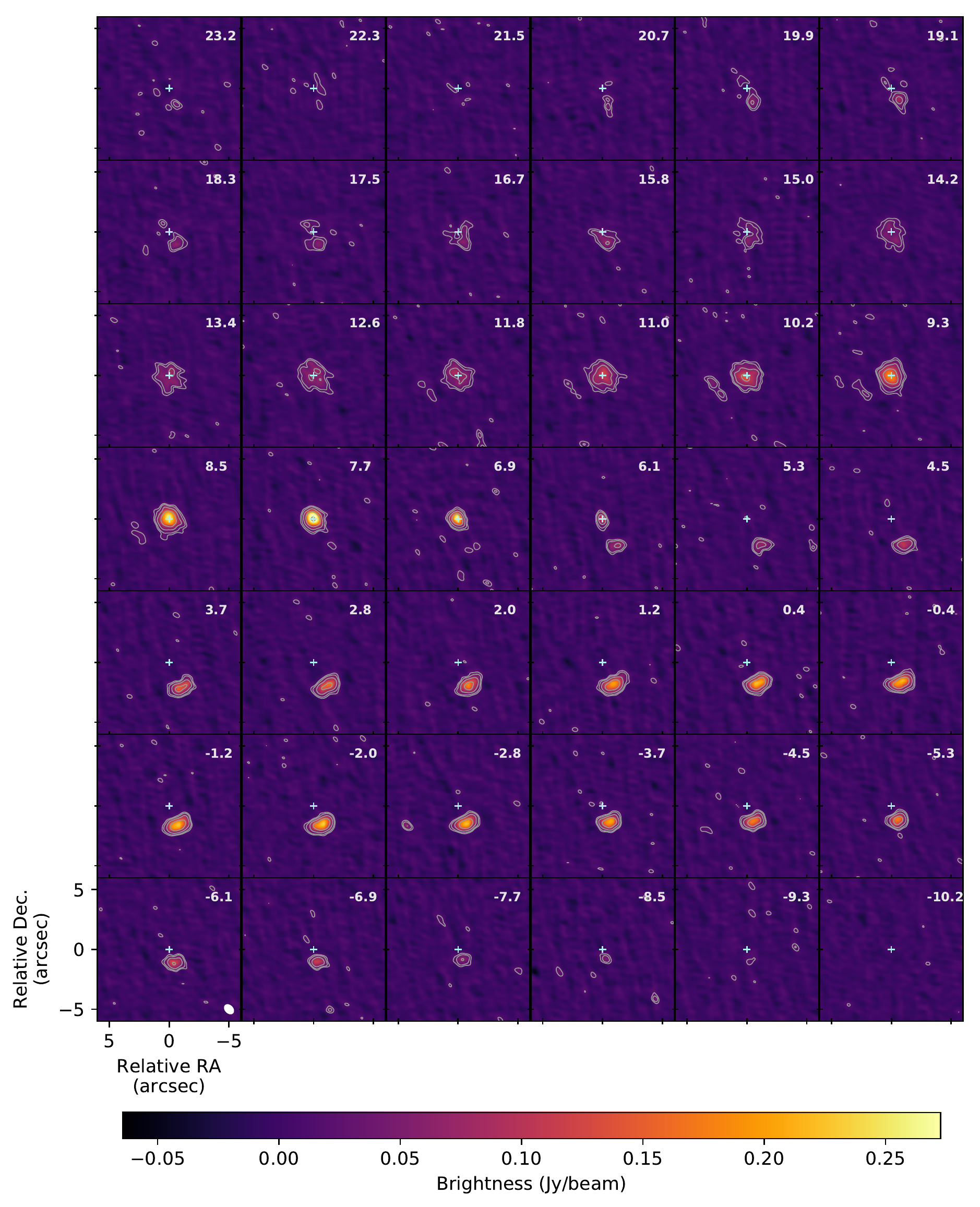}
	\contcaption{}
	\centering
\end{figure*}


\bsp	
\label{lastpage}
\end{document}